\documentclass[]{aastex631}

\newcommand{\sub}[1]{\texorpdfstring{\textsubscript{#1}}{#1}}
\newcommand{\super}[1]{\texorpdfstring{\textsuperscript{#1}}{#1}}

\shorttitle{C$_2$H in PEACHES}
\shortauthors{JGA}

\begin{document}

\title{PEACHES IV: Tracing the Formation \& Evolution of C$_2$H in Perseus Low-Mass Protostars}

\author[0009-0001-1193-2896]{Jack G. Anderson}
\affiliation{University of California, Berkeley, Berkeley CA 94720, USA}

\author[0000-0002-8716-0482]{Jennifer B. Bergner}
\affiliation{University of California, Berkeley, Berkeley CA 94720, USA}

\author[0000-0001-8227-2816]{Yao-Lun Yang}
\affiliation{RIKEN Cluster for Pioneering Research, 2-1 Hirosawa, Wako-shi, Saitama 351-0198, Japan}

\author[0000-0001-7511-0034]{Yichen Zhang}
\affiliation{Department of Astronomy, School of Physics and Astronomy, Shanghai Jiao Tong University, 800 Dongchuan Rd., Minhang, Shanghai 200240, China}
\affiliation{RIKEN Cluster for Pioneering Research, 2-1 Hirosawa, Wako-shi, Saitama 351-0198, Japan}

\author[0000-0002-3297-4497]{Nami Sakai}
\affiliation{RIKEN Cluster for Pioneering Research, 2-1 Hirosawa, Wako-shi, Saitama 351-0198, Japan}

\begin{abstract}
The radical hydrocarbon molecule C$_2$H is widely detected in various stages of star and planet formation, and has emerged as a useful tracer of high-C/O gas within the photochemically active surface layers of mature (Class II) protoplanetary disks. However, the chemistry and evolution of C$_2$H within younger (Class 0/I) protostars remains much more poorly understood.  Here, using data observed as part of the PEACHES survey along with new ALMA ACA observations, we investigate the C$_2$H emission towards an unbiased sample of 35 Class 0/I low-mass protostars in Perseus. With this large sample, we identify a clear association between C$_2$H emission and the protostellar outflow cavity walls, and a consistent spatial anti-correlation  between C$_2$H and SO emission.  Together, these trends confirm that C$_2$H is tracing photochemically active, O-poor gas in these younger sources.  We fitted the C$_2$H spectra with a simple LTE model to yield column density maps, and find values ranging from 10$^{14}$ -- 10$^{15}$ cm$^{-2}$ in these sources. We also looked for trends in the C$_2$H emission morphology as a function of various protostellar evolutionary metrics, but find no clear patterns: the C$_2$H emission remains spatially extended in most sources, independent of age.  This indicates that the transition to the compact C$_2$H emission observed on the surface of Class II disks must happen rapidly, sometime just after the embedded stage.  
\end{abstract}

\section{Introduction} \label{sec:intro}

The process of star formation is kicked off by the gravitational collapse of the densest regions of interstellar molecular clouds. Before they evolve into fully formed star systems, these objects begin the maturation process as protostars. Still embedded in their host molecular cloud, they can be characterized as pre-main sequence stars surrounded by a dense, infalling natal envelope of gas and dust (\citealt{lada1987}). In the innermost regions of these objects, material from the envelope accretes onto the newly formed star, setting its final mass and luminosity. On larger scales, this envelope begins to be dissipated by large-scale high-velocity outflows of gas and dust (\citealt{arce_sargent2006}). As these sources evolve, their outer envelope is cleared away, yielding only a host star and mature protoplanetary disk.  Between the envelope, inner envelope/proto-disk, and outflows, the protostellar environment is host to a wide range of physical conditions, which can be traced observationally using chemical probes \citep[e.g.][]{tychoniec2021}. 
\\
\par There is recent evidence that at this point in the formation of the star system, the earliest steps towards planet formation are already taking place (\citealt{harsono2018}, \citealt{manara2018}, \citealt{tobin2020}).  This makes protostars the best sites for understanding the physical processes and chemistry governing the initial formation steps of new planetary systems. Because of this connection between the formation of planets and their host stars, understanding the chemical and physical evolution of protostellar environments plays a critical role in our understanding of how planetary compositions are set inside of and beyond our solar system.
\\
\par Observations of mature Class II disks show a distinctive chemical environment compared to earlier evolutionary stages, characterized by the depletion of key volatiles in the gas and active UV-driven photochemistry (\citealt{calahan2023}). The presence of highly-reactive hydrocarbon radicals, namely C$_2$H, serves as an indicator to this unique chemical environment. Notably, strong C$_2$H emission has been widely observed towards many well studied disks (\citealt{miotello2019}, \citealt{bergner2019}), indicating a long-lived hydrocarbon chemical equilibrium (\citealt{calahan2023}). Two main factors seem to play a role in driving C$_2$H formation. Firstly, chemical models and high-resolution ALMA observations show that C$_2$H can only form efficiently in oxygen-poor, i.e. high C/O, gas (\citealt{sakai2010}, \citealt{bergin2016}, \citealt{krijt2018}, \citealt{miotello2019}, \citealt{krijt2020}, \citealt{bosman2021}, \citealt{vanclepper2022}).  This is because free gas-phase O will readily lock up gas-phase carbon into the more stable CO. Second, abundant ultraviolet (UV) radiation is required for sustained C$_2$H production (\citealt{bergin2016}, \citealt{heays2017}, \citealt{bosman2021}, \citealt{calahan2023}).  This is underscored by the early detections of C$_2$H in interstellar photodissociation regions (PDRs), which are exposed to high UV fields (\citealt{jansen1995}, \citealt{nagy2015}). The physical and chemical environments necessary for long-lived C$_2$H production in the surface layers of mature disks have been well studied, but comprehensive studies of the factors driving C$_2$H formation in younger, embedded sources are still needed. 
\\
\par C$_2$H has been detected and mapped in a number of protostellar sources including
L1527 (\citealt{sakai2014_a}, \citealt{sakai2014_b}), L483 (\citealt{oya2018}), IRAS 15398-3359 (\citealt{oya2014}, \citealt{okoda2018}), NGC 1333 IRAS 4C (\citealt{zhang2018}), and CB68 (\citealt{imai2022}). In general, the C$_2$H morphology in these embedded sources is extended.  Except for L1527 (\citealt{sakai2017}), it appears to trace the low-density cavity regions where the high velocity outflows meet the quiescent envelope in most of the sources (\citealt{tychoniec2021}). This is in contrast to the compact C$_2$H morphology seen on the surface of mature disks.  Therefore, at some point there must be a transition from the large-scale outflow morphology seen in young protostars, to the compact disk-associated morphology seen in more mature systems.  As of yet, there has not been a systematic survey of protostellar C$_2$H chemistry which could reveal this transition stage or place constraints on the timescale over which it occurs.
\\
\par With the aim of obtaining an unbiased view of C$_2$H chemistry in a large sample of protostars of varied ages, we make use of data available as part of the Perseus ALMA Chemical Survey (PEACHES, \citealt{PEACHES_I}) and the MASS Assembly of Stellar Systems and their Evolution with the SMA (MASSES, \citealt{MASSES}) which cover 35 Class 0/I protostars in Perseus. We focus on the C$_2$H, SO, and CO emission in order to trace the chemistry, morphology and evolution of C$_2$H in this sample. Our main goals are to (1) disambiguate what C$_2$H is tracing within the protostellar environment, and whether its formation is driven by then same factors (high C/O, high UV) in embedded sources as is seen in class II disks; and (2) search for evolutionary trends in the C$_2$H emission morphology, from an extended, outflow cavity structure to a more compact disk-like morphology. 

\section{Methods} \label{sec:methods}
\subsection{Observations}

\par We used ALMA observations of C$_2$H and SO transitions taken with the 12m array as part of the PEACHES project (Project codes 2016.1.01501.S and 2017.1.01462.S, \citealt{PEACHES_I}). All data reprocessing and imaging was performed using CASA (\citealt{CASA}). The measurement sets were obtained using the ALMA pipeline. Line-free channels were averaged to form continuum measurement sets, on which self-calibration was performed for sources with bright continuum emission to improve signal-to-noise ratio. We performed two to three rounds of manual self-calibration, until minimal gains in SNR were observed. The self-calibration process yielded an average improvement in continuum SNR of 31\%, while the major and minor beam axes were changed by an average of 0.2\% and 1\%, respectively. Line and continuum images were created using the CASA task \texttt{tclean}. We produced images for our spatial analysis using Briggs weighting and a robust parameter of 1. Images of 600 by 600 pixels were created, with a pixel size of 0.05". Line images were cleaned to a threshold of twice the rms estimated using line free-channels, using automasking low noise, sidelobe and noise thresholds of 1.5, 3.0 and 3.75, respectively. Moment 0 maps of the lines were then created from the line data cubes by integrating over the velocities of interest. The $N$ = 7--6, $J$ = 6--5 transition of ground state SO and  and the $N$ = 3--2, $J$ = $\frac{5}{2}$--$\frac{3}{2}$, $F$ = 3--2 C$_2$H transition were imaged. These transitions were selected for each molecule of interest because of their high signal-to-noise ratio and the overall number of sources in which detections of the transitions were made. See \citealt{PEACHES_I} for a list of all detected molecular transitions.  Additional images of S-bearing molecules towards the PEACHES sources can be found in \citet{PEACHES_III}.
\\
\par To obtain better flux measurements for column density calculations (Section \ref{sub:3}), we also used complementary observations taken with the ALMA Atacama Compact Array (ACA; project code 2022.1.00303.S) and combined with the 12m data. The ACA observations, taken in August and September of 2023, target the same PEACHES sources and molecular lines as the original 12m survey. Here only the C$_2$H spectral window is used, covering the F = 3--2, 2--1 and 2--2 hyperfine components of the N = 3--2, J = $\frac{5}{2}$--$\frac{3}{2}$ C$_2$H transition at $\sim$262.06 GHz, with a channel width of 61.035 kHz ($\sim$0.07 km/s). As before, data reprocessing and imaging was performed using CASA, and line images were created using \texttt{tclean}. The 12m and ACA data were imaged together using the `mosaic' gridding parameter, Briggs weighting and a robust value of 2. During imaging, the ACA visibilities were downweighted by a factor of 0.41 to correct for the time-on-source (ToS) ratio between the 12m and ACA observations: from 1:5.8 (actual 89:512 minutes) to an effective ratio of 1:2.4 as recommended by \citet{alma_handbook}. Line images were cleaned to a threshold of twice the rms estimated using line free-channels, using automasking low noise, sidelobe and noise thresholds of 1.5, 2.0 and 4.25, respectively. Moment 0 maps of the lines were then created from the line data cubes by integrating over the velocities of interest. 
\\
\par The CO images used in our C$_2$H spatial analyses come from SMA observations as part of the MASS Assembly of Stellar Systems and their Evolution with the SMA (MASSES) survey (\citealt{MASSES}). The MASSES survey made use of the subcompact, compact and extended configurations of the SMA, covering baseline distances of 9.5-77m, 16-77m and 44-226m, respectively. The 15-1231m baselines of the PEACHES ALMA 12m array data (configuration details provided in \citealt{PEACHES_I}), with the addition of the 9-50m baselines provided by the ACA observations, cover similar short-baseline distances.  We can therefore directly compare the large-scale emission structures from both the ALMA and SMA observations, and in addition we are able to recover more small-scale detail in the ALMA observations. CO data products were taken directly from the MASSES data release web page through the Harvard data archive (\citealt{MASSESData}).
\\
\subsection{C$_2$H Spectral Line Fitting}\label{sec:line_fitting}
\par Maps of C$_2$H column densities and rotational temperatures were produced by fitting synthetic spectra to the ACA + 12m images at positions where the F= 3--2 C$_2$H emission in the moment-0 maps is above a 3$\sigma$ threshold. The image cubes were first regridded by a factor of 4, i.e. squares of 2-by-2 pixels were averaged, to increase the signal to noise and sample on spatial scales comparable to the beam. In the spectral line model, we assume that the line profiles of each hyperfine component are Gaussian, that each hyperfine component can be described with the same rotational temperature, $T_r$, and that the emission is in local thermodynamic equilibrium (LTE).
The optical depth at the center of the each hyperfine line $i$ is found via:
\begin{equation}
    \tau_{i,0} = \frac{N_T}{Q(T_r)}e^{\frac{E_{u,i}}{T_r}}\frac{g_{u,i}A_{u,i}c^3}{8\pi\nu_i^3}\frac{1}{\sigma\sqrt{2\pi}}(e^{\frac{h\nu_i}{kT_r}}-1)
\label{eq:tau_i}
\end{equation}
where $N_T$ the total column density, $Q(T)$ the molecular partition function, and $\sigma$ is the Gaussian line width. For each hyperfine line, $E_{u,i}$ is the upper state energy, $g_{u,i}$ is the upper-state degeneracy, $A_{u,i}$ is the Einstein coefficient and $\nu_i$ is the transition frequency. The total optical depth profile can then be described as:
\begin{equation}
    \tau_\nu=\sum_{i} \tau_{i,0}\;\mathrm{exp}\left( \frac{-(V-V_i-V_{lsr})^2}{2\sigma^2}\right)
\label{eq:tau_nu}
\end{equation}
where $V$ is the velocity, $V_i$ is the velocity offset of each hyperfine line, and $V_{lsr}$ is the source velocity. Finally, the intensity can be calculated using:
\begin{equation}
I_{\nu} = [B_\nu(T_r) - B_\nu(T_{bg})]\times(1-e^{-\tau_\nu})\times\Omega
\label{I_nu} 
\end{equation}
where $B_\nu$ is the Planck function, $\Omega$ is the effective angular area of the restoring beam, and $T_{bg}$ the cosmic microwave background (2.73 K).

\section{Results} \label{sec:results}

\subsection{C$_2$H and SO Emission Morphology} \label{sub:1}

\begin{figure}[h!]
    \centering
    \includegraphics[width=6.25in]{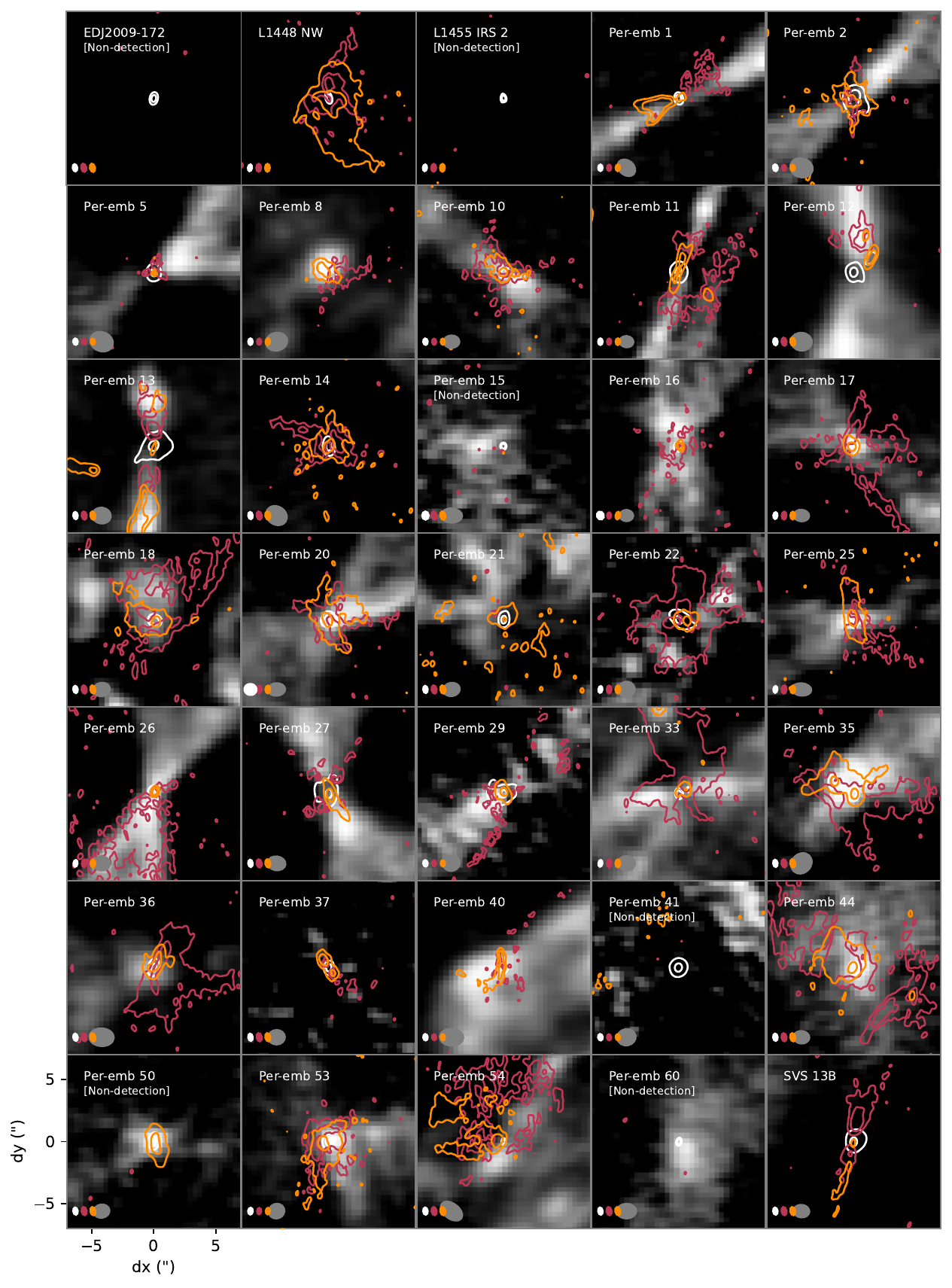}
    \caption{Perseus protostellar sources with continuum detections. Integrated intensity maps of C$_2$H and SO are contoured in red and orange, respectively. White contours show the mm continuum. Grey colormaps correspond to both the red and blueshifted CO (2-1) emission. Synthesized beams for all maps are provided in the bottom left of each panel.}
    \label{fig:fig1}
    \end{figure}

Previous observations of C$_2$H in protostars have hinted at a connection between the C$_2$H emission and the outflow (\citealt{oya2014}, \citealt{oya2018}, \citealt{tychoniec2021}), as well as a possible anti-correlation with SO (\citealt{sakai2014_a}, \citealt{sakai2014_b}, \citealt{okoda2018}, \citealt{arturdelavillarmois2019}). Using the unbiased survey of C$_2$H provided by the PEACHES dataset, we aim to explore these connections by comparing the C$_2$H morphology to the outflow tracer CO, as well as SO. In studying these correlations, we are interested in the small-scale emission patterns, and therefore use the 12m-only images for our analysis.
Figure \ref{fig:fig1} presents an overview of the PEACHES sources and detected molecules of interest. The two sources where continuum emission is not detected, SVS 3 and L1448 IRS2E, are omitted. Integrated intensity (moment-0) maps were constructed for all the sources where C$_2$H emission was detected, shown contoured in red. Also included in Figure \ref{fig:fig1} are integrated maps of SO, contoured in orange. Integrated intensity maps of CO are shown as a greyscale colormap, which includes both red- and blueshifted emission.

    \clearpage
\par Of the 35 protostars observed, C$_2$H and SO were detected towards 30 and 30, respectively. The sources shown in Figure \ref{fig:fig1} display a wide range of C$_2$H emission morphologies. Some appear relatively extended (scales of $\sim$3-4") but without discernible structure, (see L1448 NW, Per-emb 18, Per-emb 33, Per-emb 36, Per-emb 44, Per-emb 54). However, many sources exhibit coherent C$_2$H emission morphologies that suggest a relationship to the outflow structures traced by CO. Several of these exhibit quite narrow C$_2$H emission (Per-emb 11, 13, SVS 13B), while most others exhibit a X-wing or cone-like emission morphology (Per-emb 1, 2, 5, 10, 14, 17, 20, 22, 25, 26, 27, 29, 35, 53; and to a lesser extent 8 and 12). Interestingly, many of the sources with this cone-like structure are also those where the CO (2-1) moment-0 maps (in greyscale) trace discernible bipolar outflows moving away from the continuum peak.  In these cases, we see a clear association between the C$_2$H emission and the base or edges of the outflow traced by CO.
\\
\par
\begin{figure}[h!]
    \centering
    \includegraphics[width=6in]{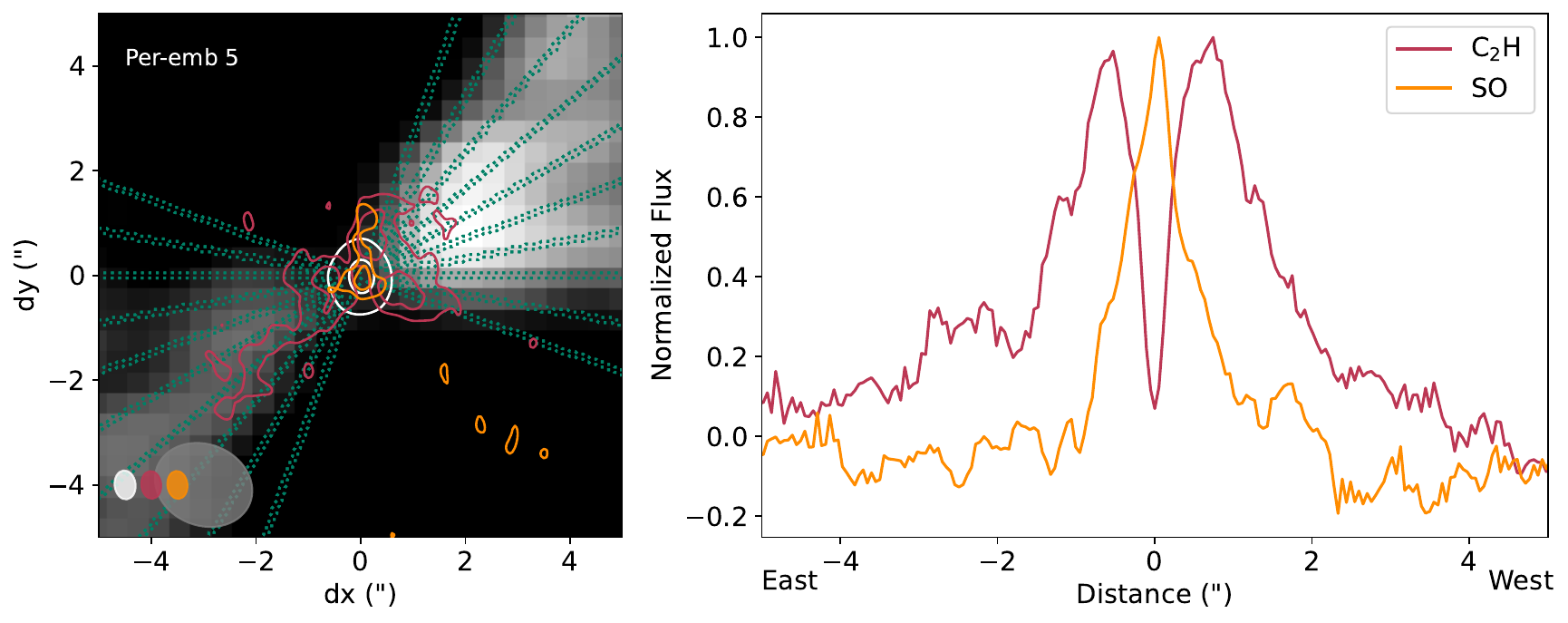}
    \caption{Example spatial analysis of the C$_2$H and SO emission. Left: red and orange contours correspond to the respective C$_2$H and SO integrated intensity maps. The grey colormap and white contours in the left image correspond to the CO (2-1) integrated intensity map and mm continuum, respectively. Green lines show the rays along which the C$_2$H and SO intensity profiles are sampled.  Right: spatial intensity profile through the outflow region, made by averaging the profiles along the individual rays.  A distance of 0$\arcsec$ represents the continuum peak.}
    \label{fig:fig2}
    \end{figure}

 
The SO moment-zero maps (contoured in orange) show a similar range of behavior to that of C$_2$H, with some directionally extended emission and some more compact emission.  However, comparison of SO and C$_2$H by visual inspection of the moment-zero contours is not sufficient to understand the degree of spatial overlap between the two molecules, especially given that many sources show large azimuthal asymmetries in their emission patterns.
To more definitively compare the C$_2$H and SO emission morphology in the outflowing regions of the protostars, we produce a series of spatial intensity profiles, as shown for an example source in Figure \ref{fig:fig2}. The emission of both molecules of interest is sampled along spatial rays (shown in green) which approximately cover the outflow regions around each protostar.  The rays are spaced in 10 degree increments originating at the source continuum peak. The emission from each individual ray is then averaged to give a spatial intensity profile, as shown on the right of Figure \ref{fig:fig2}.

As shown in Figure \ref{fig:fig2} for the case of Per-emb 5, this procedure clearly illustrates the difference in spatial morphology between the two molecules. The C$_2$H is more spatially extended, with emission extending up to 2$\arcsec$ away from the continuum peak. The SO, as can be visualized in both the moment-zero map and intensity profile, is compact around the location of the continuum maximum, only showing emission within $\sim$1$\arcsec$ around the continuum peak. What is not clear from the moment-zero contours, but does become evident from the average intensity profiles, is the drop in C$_2$H emission towards the continuum peak of Per-emb 5, as shown by the dip in the C$_2$H intensity profile around a distance of 0$\arcsec$.  This coincides with the SO emission peak.  Moving away from the continuum peak, the C$_2$H emission climbs sharply while the SO emission falls off. This type of behavior is very similar to the profile observed in IRAS 15398-3359 (\citealt{okoda2018}). To analyze this further, this type of intensity profile analysis was carried out across all sources in the sample where a bipolar outflow could be discerned from the CO (2-1) emission.

\clearpage

    \begin{figure}[hb!]
    \centering
    \includegraphics[width=6in]{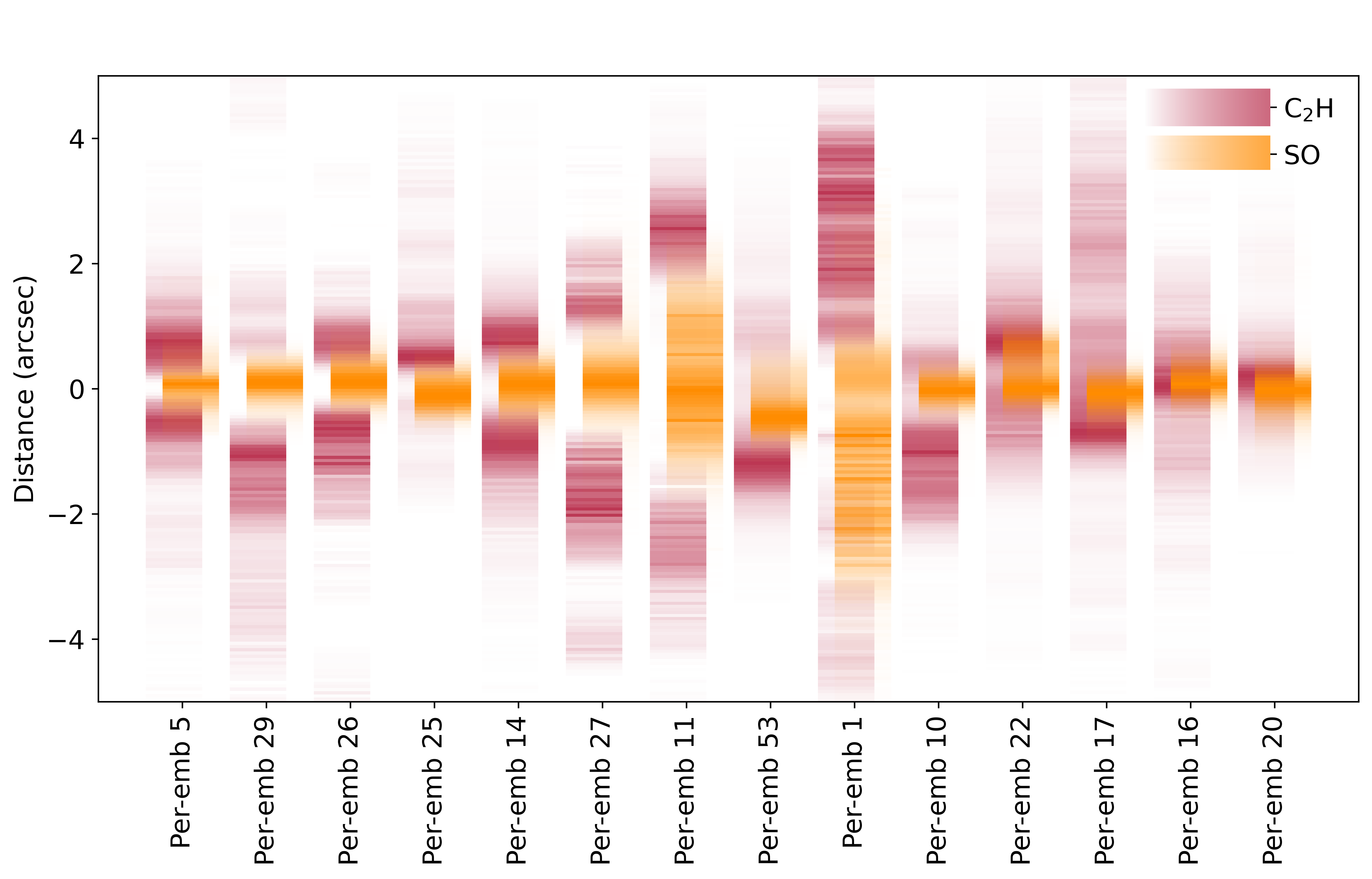}
    \caption{Average intensity profiles of C$_2$H (red) and SO (orange) for the sources with discernible bipolar outflows. Intensity profiles are sampled only within the outflow region (Figure \ref{fig:fig2}), and distances away from the continuum peak for each source are shown on the y axis.  The color transparency represents the normalized flux at each spatial position, with the highest saturation representing the maximum flux across the spatial profile.}
    \label{fig:fig3}
    \end{figure}

Figure \ref{fig:fig3} shows the resulting intensity profiles in all sources for which this spatial analysis was possible. To enable comparisons across the sources, the average intensity profiles are now plotted with distance as the vertical axis and the relative emission intensity of each molecule shown with color saturation. The sources have been arranged such that similar SO and C$_2$H morphologies are grouped together. The leftmost sources, Per-emb 5, 29, 26, 25 and 14 all display compact SO around the continuum peak, whereas C$_2$H is absent from the continuum peak but emits at larger distances beyond the SO peak.  Per-emb 10, 22, and 17 also show compact and centrally located SO, with C$_2$H emission dipping at the SO position but not disappearing entirely. In the center sources (Per-emb 27, 11, 53 and 1), the SO emission is either spatially extended or off-center from the continuum peak; even so the C$_2$H emission again anti-correlates with the SO emission locations. Per-emb 1 is particularly asymmetric, with strong SO emission within one outflow lobe and strong C$_2$H emission in the other lobe. This can also be visualized in the moment-zero maps of Per-emb 1 presented in Figure \ref{fig:fig1}. The sources on the right, Per-emb 16 and 20, exhibit compact SO emission, but the C$_2$H emission does not drop around the the continuum peak and appears more co-spatial with SO.  Overall, we find that most sources (12 of 14) show a clear anti-correlation between the location of C$_2$H and SO within the outflow.

\subsection{ACA Images and Column Density Maps}\label{sub:3}

\begin{figure}[htb!]
    \centering
    \includegraphics[width=6in]{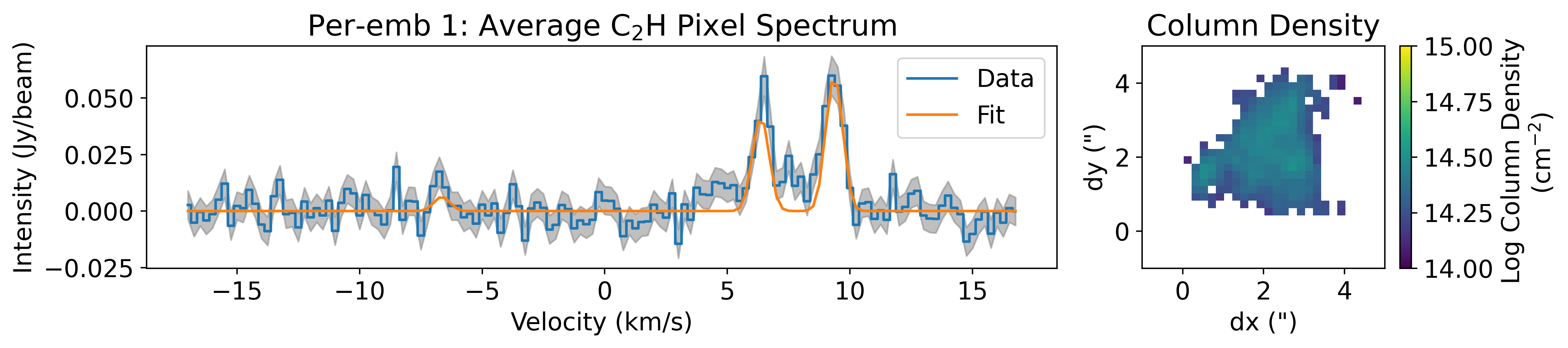}
    \includegraphics[width=6in]{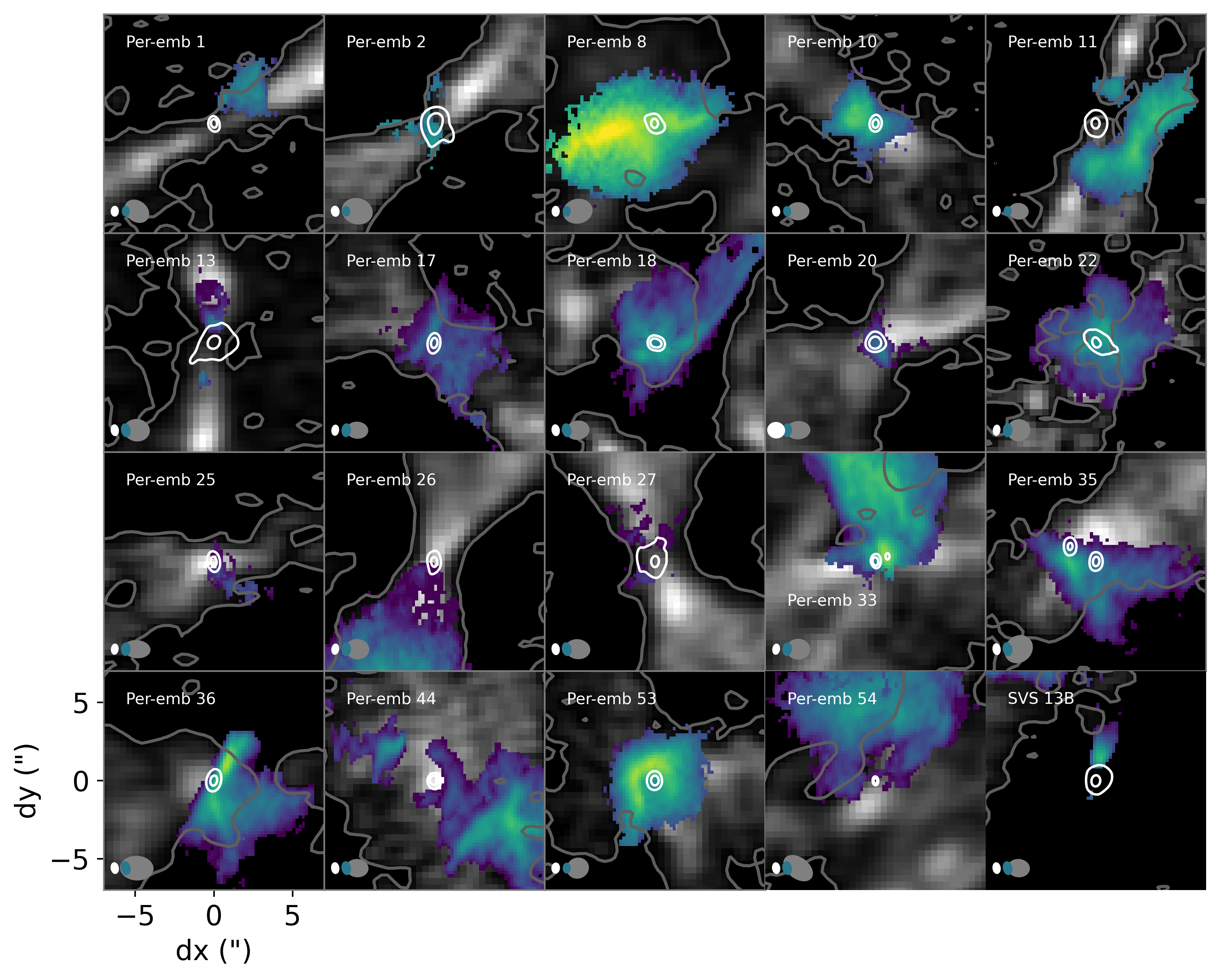}
    \caption{Fitted C$_2$H column density maps using combined ACA and 12m images. Top row: example spectral line model fitted to the average C$_2$H spectrum in Per-emb 1 (left, data error plotted in grey), with map of retrieved column density (right). Bottom panel: Fitted column density maps of sources with bright C$_2$H emission, beam sizes shown in blue. CO (2-1) emission (greyscale colormap with grey contours) and the mm continuum (white contours) are also shown for reference. All column density colormaps have the same color scaling as the example of Per-emb 1.}
    \label{fig:fig5}
    \end{figure}

Beyond the characterizing the emission morphology, we also aim to obtain quantitative constraints on the C$_2$H column densities towards the sources. To do this, we create column density maps using the spectral line model outlined in section \ref{sec:line_fitting}. To ensure complete recovery of the C$_2$H flux, we use the combined 12m+ACA images for this analysis. To illustrate our approach, the top panel of Figure \ref{fig:fig5} shows the average spectrum for Per-emb 1 (from all pixels with a $>$3$\sigma$ detection of the F=3-2 hyperfine component), with the fitted model overlaid in orange. Our C\sub{2}H hyperfine transitions span an equal upper state energy, and as a result cannot precisely fit a variable rotational temperature. We proceed with fitting column densities assuming a fixed temperature of 20K, and bound our column density values between 10\super{13} and 10\super{16} cm\super{-2}, noting that true constraints on the C$_2$H rotational temperatures, and in turn more precise constraints on the column densities, would require coverage of additional C$_2$H lines. Our choice of rotational temperature is informed by prior studies of molecular outflows in low-mass protostars, from which rotational temperatures of 10--50K are typical (\citealt{Parker1991}, \citealt{Hatchell2007}, \citealt{Curtis2010}, \citealt{Dunham2010}, \citealt{Dunham2014}). We adopted a value on the low end of this range under the assumption that the outflow cavities are cooler than the outflowing gas itself. In any case, varying T$_r$ in our spectral line modeling within the 10--50 K range yields changes in calculated C$_2$H column densities of $<$ 5\%. 
\\
\par
The maps resulting from fitting all pixels individually are shown in the top panel on the right. Column density maps for all sources with sufficiently bright C$_2$H emission are shown in the bottom portion of Figure \ref{fig:fig5}.  For reference, we also show the CO 2-1 emission (grey colormap) and mm continuum (white contour). All column density colormaps in the lower panel have the same color scaling as the example of Per-emb 1. The fitted column density maps show values spanning $\sim 10^{14}$ -- $10^{15}$ cm\super{-2}, with maps resembling that of the flux, as expected when fitting assuming a fixed rotational temperature. See Table \ref{tab:column densities} for the median and range of fitted column densities for all sources. The average-per-source uncertainties on our retrieved column densities range from 10\super{0.05} - 10\super{0.5} cm\super{-2}. Most sources show high C\sub{2}H column densities close to the continuum max, decreasing moving away form the central star, while some (Per-emb 26, 44, and 54) show peaking C\sub{2}H column densities more offset from the continuum max. However, definitive conclusions would require maps of fitted column densities with variable rotational temperatures, only possible with further spectral coverage.  

\clearpage

\subsection{C$_2$H Morphology in Conjunction with Protostar Evolution}\label{sub:2}

\par Using this large sample of protostellar sources with C$_2$H detections, we also attempt to identify possible evolutionary trends in the C$_2$H emission over the course of the protostellar lifetime. In order to do this, we compare the C$_2$H emission with three evolutionary metrics. As embedded protostars evolve, their central dust disk, best traced in the mm continuum, is expected to increase in size \citep[e.g.][]{Maury2019}.  Additionally, their observed luminosities increase and their spectral energy distributions (SEDs) shift towards longer wavelengths (\citealt{dunham2014review}). This shift in SED can be described using a blackbody with the same mean frequency, whose temperature increases as the SED shifts towards longer wavelengths. This temperature is defined as the bolometric temperature, and luminosity described as the bolometric luminosity (\citealt{myers1993}, \citealt{Myers1998}). We therefore use the mm continuum deconvolved radius, the bolometric temperature and the bolometric luminosity as our evolutionary proxies. In particular, we are interested to see whether the C$_2$H emission trends from more extended morphologies associated with large-scale outflow cavities to more compact morphologies associated with a (proto-)disk over the duration of the embedded stage.
\\
\par
\begin{figure}[h!]
    \centering
    \includegraphics[width=6in]{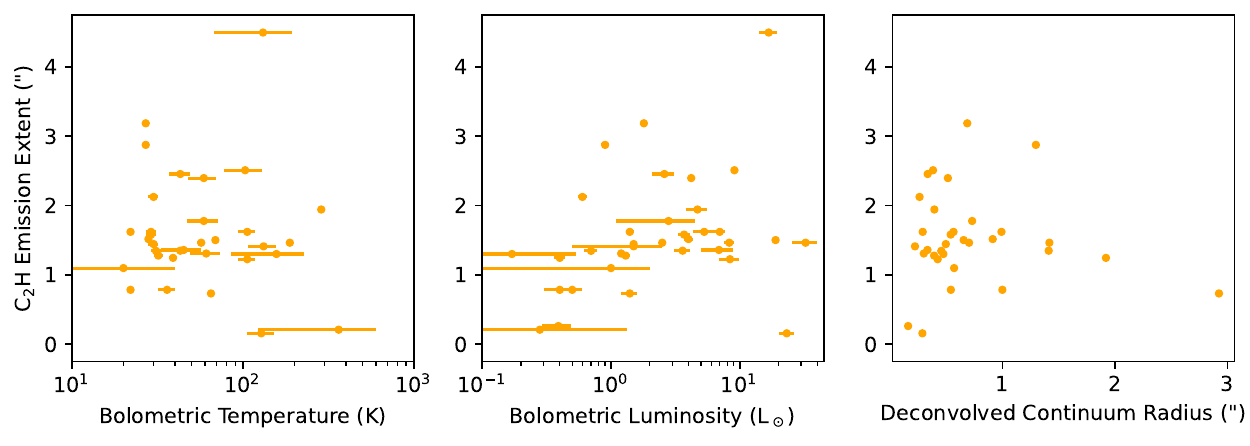}
    \caption{C$_2$H emission extent as a function of protostellar evolution. The width above 3$\sigma$ of the C$_2$H spatial intensity profile is plotted with respect to the three evolutionary metrics: bolometric temperature and luminosity (\citealt{tobin2016}; uncertainties shown with errorbars) and mm deconvolved radius. For sources where bipolar outflows could not be identified, the intensity profiles are sampled along all directions from the continuum.}
    \label{fig:fig4}
    \end{figure}

 For this comparison, we estimated the overall spatial extent of the C$_2$H emission from the width above 3$\sigma$ of the intensity profiles obtained in Section \ref{sub:1}. These are plotted in Figure \ref{fig:fig4} against the three evolutionary metrics. The bolometric temperature and luminosity values and uncertainties are taken from \citealt{tobin2016}, while the deconvolved radii are determined via the CASA task \texttt{imfit}. The sources clearly vary in spatial extent of the C$_2$H emission, however there is no clear discernable trend with respect to our evolutionary metrics. The C$_2$H emission may be either extended or compact for both less mature and more mature systems.


\clearpage

\subsection{Kinematic Comparison of C\sub{2}H with CO}

\par
\begin{figure}[htb!]
    \centering
    \includegraphics[width=\textwidth]{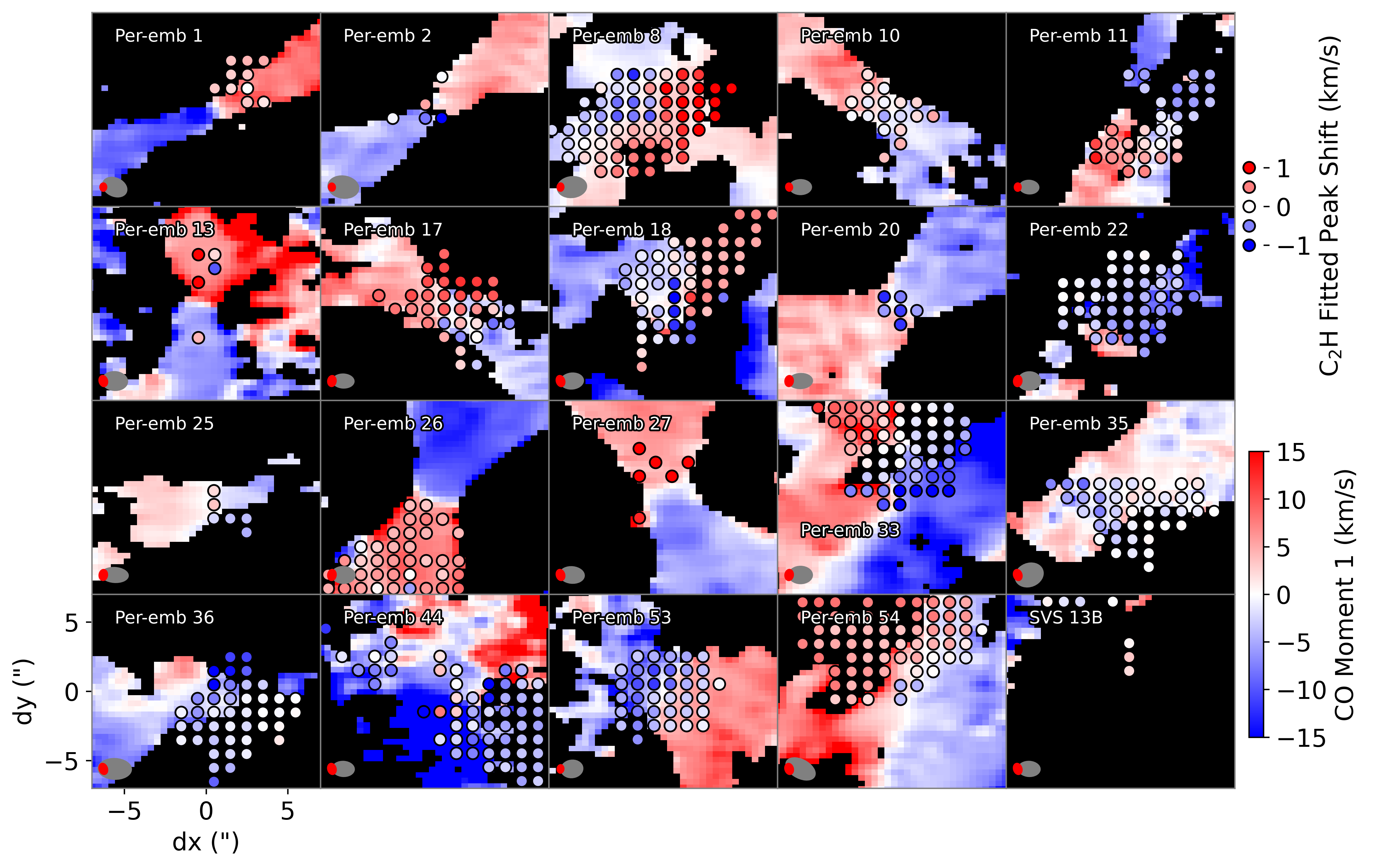}
    \caption{Kinematic comparisons between C\sub{2}H and CO. Moment-1 maps of the CO are shown as a colormap, while the coloring of the C\sub{2}H circles correspond to the shift in velocity of the fitted F=3-2 hyperfine component, based on the spectral fitting done in section \ref{sub:3}. The grey and blue ellipses in the bottom corner of each image correspond to beam sizes of the CO and C\sub{2}H observations, respectively.} 
    \label{fig:C2H_CO}
    \end{figure}

 We also make use of the C\sub{2}H fitting procedure outlined in Section \ref{sec:line_fitting} to extract basic kinematic information from the observations and compare that to the outflows as traced by CO. From our fitted spectra we calculate the shift in the C$_2$H F=3-2 hyperfine component with respect to the source velocity, and compare that to the intensity-weighted velocities (moment-1) of the CO observations. These maps are shown in Figure \ref{fig:C2H_CO}. Most sources show some red- or blue-shifting of the C\sub{2}H emission, though with a fairly small magnitude of $\sim$ 1 km/s, still much larger than the 0.07 km/s velocity resolution of our observations. 
\\
\par In Per-emb 1, 10, 11, 17 and 26, the C\sub{2}H emission most closely traces bipolar outflow cavities (Section \ref{sub:1}). Of these, the C$_2$H in Per-emb 10 appears very close to the rest velocity. Per-emb 11 and 17 show gradients in the C\sub{2}H kinematics that appear correlated with the kinematic structure of CO, with a transition from red-to-blueshifted emission across the two lobes of the outflow. Interestingly, this pattern holds for both of the outflow-like structures near Per-emb 11. In Per-emb 1, 26, and 27, C$_2$H only traces one outflow lobe, but in both cases shows redshifted emission consistent with the CO kinematics. Per-emb 33 also shows similar kinematic trends between C$_2$H and CO, but has less clearly defined outflow structure.  Note that in all cases, the magnitude of the C$_2$H red-or blue-shifting is much lower than the CO velocities.  Still, these correlations suggest some connection in the velocity structure between the gas of the outflow cavity and that of the outflow itself.  In several sources where no bipolar outflow is clearly identifiable, the C$_2$H emission is uniformly blueshifted (Per-emb 22, 35, 36, 53), possibly due to the extinction of emission originating on the far side of the protostar.  
\\
\par
It is important to consider whether radiative transfer effects may be influencing our comparison between the CO and C$_2$H kinematics.  We appear to be tracing optically thinner, higher-velocity CO from the inner region of the outflows, rather than optically thick CO which would originate from the outflow's outer layers at a lower velocity shift.  Therefore, we do not expect that optical depth effects are responsible for any differences in red/blueshift between the two species.

\clearpage


\section{Discussion} \label{sec:discussion}

\subsection{Evolution of C$_2$H chemistry}
This large sample of low-mass protostars provides an opportunity to search for evolutionary patterns in the emission morphology of C$_2$H.  As protostars begin to clear away their envelope and mature towards Class II systems, the photochemically active regions could be expected to evolve \citep[e.g.][]{bergner2020}.  In particular, at some stage before $\sim$1 Myr, photochemistry must become active on small scales at the disk position based on the C$_2$H morphologies seen in Class II disks (\citealt{bergin2016}, \citealt{bergner2019}, \citealt{guzman2021}).
Figure \ref{fig:fig4} shows, however, that there is no clear correlation between the C$_2$H emission extent in these Class 0/I sources and the evolutionary metrics of bolometric temperature/luminosity or continuum radius. This can also be seen through visual inspection of Figure \ref{fig:fig1}, in which the sources are ordered in terms of increasing bolometric temperature (or increasing maturity).  While there are some systems with more compact C$_2$H emission confined to the outflow base (Per-emb 2, 5, 10, 16, 20, 27), which could approximately correspond to a proto-disk surface, there is no obvious evolutionary association.  One caveat is that we cannot rule out whether there is a compact C$_2$H component on even smaller spatial scales than we can access with the spatial resolution of these observations.
\\
\par
The fact that the C$_2$H emission morphology does not apparently evolve during the Class 0/I lifetime implies that a rapid change must happen at the Class I/Class II transition.  The detection of bright, compact C$_2$H emission even in relatively young $\lesssim$1 Myr Class II disks \citep{miotello2019, bergner2019} means that small-scale photochemistry must `turn on' soon after the embedded protostar phase. This fast photochemical transition is one of several dramatic chemical changes that seems to occur during the transition from Class I to Class II systems, along with the rapid depletion of CO (\citealt{zhang2020}, \citealt{bergner2020}) and H\sub{2}O (\citealt{harsono2020}).  Studying the volatile chemistry in intermediary sources between the embedded protostellar and mature disk stages will be crucial to disentangling these processes.

\subsection{C$_2$H formation chemistry in embedded protostars}
Our study of C$_2$H in Perseus protostars provides insight into the underlying C$_2$H chemistry in these embedded environments.  In mature systems, C$_2$H is associated with the ingredients of high-C/O gas, and high UV fields \citep{bergin2016}.  We show here that C$_2$H can be used to trace both of these conditions in the embedded stage as well.

The zeroth-moment maps in Figure \ref{fig:fig1} reveal that many of the Perseus protostars exhibit C$_2$H emission with a motif of x-wing or cone like structure, which widen moving away from the continuum. The coincidence of these directional, cone or wing-like structures with the edges of the CO emission strongly points to a relationship between C$_2$H and the overall outflow structure. With this large and unbiased sample, we can confirm the trend between C$_2$H and outflow cavity walls previously identified in individual protostars: B1-c \citep{tychoniec2021}, CB68 \citep{imai2022}, and IRAS 15398–3359 \citep{oya2014,okoda2018}. 
The presence of C$_2$H and similar hydrocarbons (e.g.~c-C$_3$H$_2$ in SMM3; \citealt{tychoniec2021}) at the intersection between the outflow and the envelope can be attributed to the prevalence of UV photons in these low density regions, which are needed to produce hydrocarbon radicals. The fact that several protostars do not clearly show this association or it appears incomplete may be due to unfavorable viewing geometries (e.g.~if the outflow is pointing toward or away from the line of sight), more complicated source structures or the chemical effects discussed below. Indeed, in all cases where we could clearly identify a bipolar CO outflow structure, we see C$_2$H emission associated with the base and/or edges of at least one side of the outflow cavity.
\\
\par We can also investigate the role of gas-phase oxygen in suppressing C$_2$H production by comparing the C$_2$H and SO emission morphologies.  SO is associated with the sputtering of icy material from grains and therefore serves as a signpost of oxygen rich gas (\citealt{bachiller1997}, \citealt{Jimenez-Serra2005}).  Figure \ref{fig:fig3} shows a clear spatial anti-correlation between C$_2$H and SO, implying that C$_2$H production cannot be sustained in gas that is oxygen rich.
Indeed, of the 14 Perseus sources where C$_2$H is present and the outflow could be clearly identified, 12 sources exhibit strong avoidance between SO and C$_2$H. This behavior in sources where the SO is compact and co-spatial with the continuum (Per-emb 5, 29, 26, 25) could be attributed to differing physical or chemical factors close to the protostar.  However, this cannot explain the avoidance of the two species in the sources where the SO emission is offset from the continuum (Per-emb 27, 14, 11, 53, 1).  The fact that this trend exists for a range of SO and C$_2$H structures points to the destruction of C$_2$H in regions traced by SO.
\\
\par
This behavior is consistent with expectations from chemical modeling: in oxygen rich gas, available gas-phase carbon will react with oxygen or O-bearing species, thereby suppressing the formation of hydrocarbons \citep[e.g.][]{Bergin1997, Hollenbach2009, Du2015}. \citealt{bosman2021} has shown this threshold gas phase C/O ratio is dependent of the UV field strength and overall gas density, but for a UV field to gas density ratio ($F_{UV}/n_{gas}$) of 10\super{-7} - 10\super{-5} G\sub{0} cm\super{3}, this hydrocarbon production is quenched as the C/O falls below unity. Interestingly, spatial anti-correlations between small hydrocarbons and SO are also seen in the mature Herbig disks HD 169142 and HD 100546 \citep{booth2023,booth2024}.  C$_2$H can therefore serve as a reliable tracer to distinguish O-rich vs. O-poor gas within protostellar environments.  Indeed, while the SO observed in our sources is likely produced at least in part by weak-shock sputtering of grains (\citealt{PEACHES_II}), we expect O-rich gas to be present on even smaller spatial scales due to the thermal sublimation of icy material, for instance the spatially unresolved detections of O-bearing complex organics reported in \citet{PEACHES_I}.  Higher-resolution observations of organics as well as C$_2$H would provide a more detailed picture of the balance between O- and C-dominated gas on smaller, disk-forming scales in the protostellar environment.
\\
\par
While the C$_2$H morphologies are quite different between embedded protostars and mature disks, the column densities are relatively consistent.  As shown in Figure \ref{fig:fig5}, we retrieve column densities ranging from 10\super{14} - 10\super{15} cm\super{-2} with an average C$_2$H column density across all sources of 10\super{14.3 $\pm$ 0.2} cm\super{-2}. This is comparable to values estimated in the Class II sources in \citealt{guzman2021}, and to the column densities predicted by PDR chemical models \citep{nagy2015}.  The similar column densities across these varied contexts points to a formation chemistry dependent on a narrow range of physical conditions: i.e., if C$_2$H is always formed within a thin gas layer characterized by high UV fluxes, then the resulting column density may be similar even in different astrophysical objects. Note, however, that coverage of additional J-level transitions would improve the precision of our column density retrievals.


\section{Summary} \label{sec:summary}
Using ALMA 12m and SMA data as part of the PEACHES and MASSES projects and new ALMA ACA observations, we analyze the C$_2$H emission morphology towards 35 embedded protostars in Perseus.
\begin{itemize}
    \item We present zeroth-moment maps of the C$_2$H in the the sources, and show many exhibit a emission morphology which traces the protostar outflow cavity.
    \item Analysis of the C$_2$H and SO emission morphologies though the use of outflow intensity traces shows strong spatial avoidance between the two molecules, demonstrating the role of gas-phase oxygen in the formation chemistry of C$_2$H.
    \item The C$_2$H emission morphology appears to not be directly related to protostar age though comparisons of C$_2$H structure with source bolometric temperature and deconvolved continuum radius, however all sources where C$_2$H is detected exhibit large, extended structure, indicating the transition in C$_2$H morphology to what is seen in mature disks must happen rapidly.
    \item The C$_2$H spectra of combined 12m and ACA images is fit to a spectral line model, and maps of the fitted column density show C$_2$H column densities of ~10\super{14} - 10\super{15} cm\super{-2}.
\end{itemize}
Through these observations and analysis, we are able to better understand C$_2$H morphology and the role of oxygen in the formation of C$_2$H over a large survey of protostellar sources. Additionally, we better understand the rapid transition that these low-mass sources must undergo from the embedded stage to the mature disk stage, as the photochemically active regions that C$_2$H trace go from extended and surrounding the outflows to on the surface of the disk once the star is clear of its envelope. To further investigate the difference between the C$_2$H distribution around Class 0/I sources and that in disks around Class II sources, analyzing the correlation between the outflow opening-angle and the distributions of C$_2$H and outflow-tracing molecules may serve as a useful diagnostic.  Additionally, in some sources with high C$_2$H abundances, other mechanisms beyond C$_2$H formation in the cavity wall may need to be considered \citep[e.g. Warm Carbon-Chain Chemistry triggered by CH$_4$ evaporation][]{sakai2013, aikawa2020}. Further studies of C$_2$H in sources just at the end of their embedded protostar stage may yield more information about when in planet formation period C$_2$H and other photochemically active molecules may be formed in regions undergoing the very first stages of planet formation.

\section{Acknowledgments}
The PEACHES project is supported by the MEXT/JSPS Grant-in-Aid from the Ministry of Education, Culture, Sports, Science, and Technology of Japan (JP20H05844 and JP20H05845).  J.B. and J.A. are grateful for support from UC Berkeley and the UC Berkeley College of Chemistry.  This research used the Savio computational cluster resource provided by the Berkeley Research Computing program at the University of California, Berkeley (supported by the UC Berkeley Chancellor, Vice Chancellor for Research, and Chief Information Officer).  Y.-L.Y., Y.Z., and N.S. are grateful for support from the RIKEN pioneering project: Evolution of Matter in the Universe.  This paper makes use of the following ALMA data: ADS/JAO.ALMA\#2016.1.01501.S, ADS/JAO.ALMA\#2017.1.01462.S., and ADS/JAO.ALMA\#2022.1.00303.S. ALMA is a partnership of the ESO (representing its member states), the NSF (USA) and NINS (Japan), together with the NRC (Canada) and the NSC and ASIAA (Taiwan), in cooperation with the Republic of Chile. The Joint ALMA Observatory is operated by the ESO, the AUI/NRAO, and the NAOJ.\\

Facility: ALMA

\clearpage
\section{Appendix: 12m + ACA Images}
\begin{figure}[h!]
    \centering
    \includegraphics[width=5.5in]{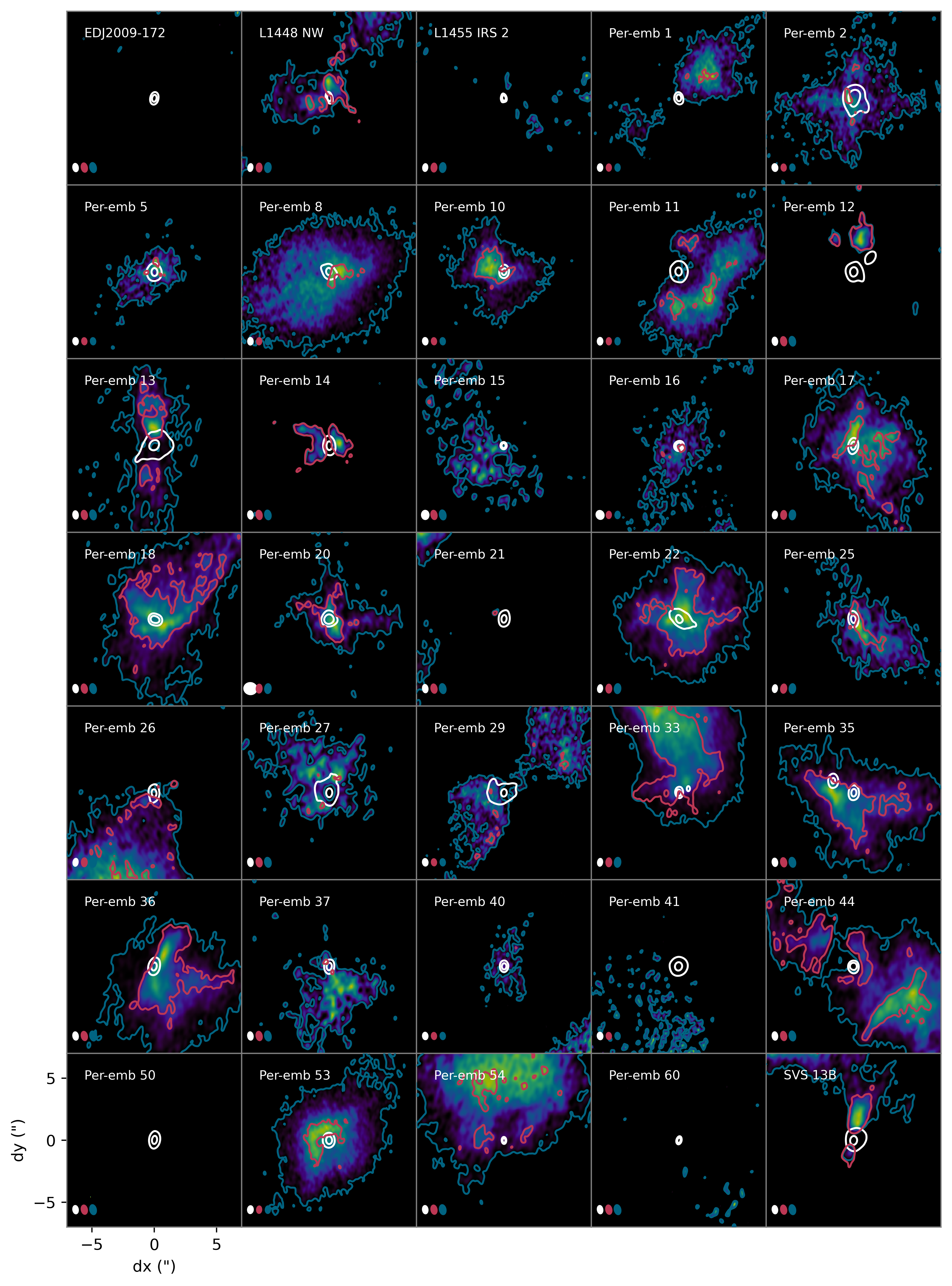}
    \caption{Combined 12m + ACA and 12m C$_2$H moment-0 maps. 12m moment-0 maps are contoured in red, and the combined 12m + ACA moment-0 maps are contoured in blue and shown as colormaps. The colormap minmum, and all contours are drawn at 35 mJy/beam for all sources.}
    \label{fig:fig6}
    \end{figure}

In Figure \ref{fig:fig6}, we present the moment-0 maps of the C$_2$H emission from both the 12m ALMA PEACHES data as well as the combined 12m + ACA images. The red contours show the 12m only and the blue contours and colormap show the 12m + ACA images. For all sources, the contours are drawn at 35 mJy/beam as well as the colormap minimum. The combined 12m + ACA images, as compared to the imaging the 12m data alone contoured at the same flux of 35 mJy/beam reveals more large scale structure in the C$_2$H surrounding the sources, with the exception of Per-emb 14. As expected, these images still retain the general spatial morphology seen in the 12m images. As the combined 12m + ACA images are less sensitive to small scale structure, we use the 12m only images for the spatial analyses with SO in section \ref{sub:1}.

\section{Appendix: Line and Continuum Imaging Tables}

\begin{table}[h!]
\centering
\caption{Continuum Parameters}
\begin{tabular}{c|ccc}
\hline
Source & Max (mJy/beam) & RMS (mJy/beam) & Beam (") \\
\hline
Per-emb 53 & 18.30 & 0.09 & 0.59 x 0.42 \\
Per-emb 16 & 6.00 & 0.13 & 0.69 x 0.59 \\
Per-emb 8 & 56.86 & 0.14 & 0.57 x 0.39 \\
Per-emb 11 A & 127.53 & 0.29 & 0.58 x 0.40 \\
Per-emb 1 & 53.60 & 0.62 & 0.57 x 0.39 \\
Per-emb 5 & 169.17 & 0.31 & 0.56 x 0.40 \\
Per-emb 2 & 115.35 & 0.54 & 0.59 x 0.43 \\
Per-emb 40 & 13.58 & 0.09 & 0.56 x 0.40 \\
Per-emb 10 & 18.15 & 0.09 & 0.56 x 0.40 \\
Per-emb 29 & 95.14 & 0.25 & 0.56 x 0.40 \\
Per-emb-41 N/S & 128.81 & 0.25 & 0.56 x 0.41 \\
L1448 NW & 56.25 & 0.71 & 0.59 x 0.36 \\
Per-emb 33 A & 155.65 & 0.39 & 0.59 x 0.36 \\
Per-emb 26 & 145.36 & 0.48 & 0.60 x 0.37 \\
Per-emb 22 A & 30.46 & 0.17 & 0.59 x 0.36 \\

Per-emb 25 & 86.84 & 0.22 & 0.59 x 0.37 \\
Per-emb 17 & 34.07 & 0.17 & 0.59 x 0.37 \\
Per-emb 20 & 8.25 & 0.38 & 1.01 x 0.93 \\
L1455 IRS 2 & 2.04 & 0.15 & 0.58 x 0.37 \\
Per-emb 44 & 148.83 & 2.13 & 0.63 x 0.40 \\
Per-emb 12 & 502.33 & 1.06 & 0.63 x 0.40 \\
Per-emb 13 & 311.00 & 0.51 & 0.63 x 0.40 \\
Per-emb 27 & 141.23 & 0.34 & 0.63 x 0.40 \\
Per-emb 54 & 1.62 & 0.09 & 0.63 x 0.40 \\
Per-emb 21 & 43.37 & 0.16 & 0.63 x 0.40 \\
Per-emb 14 & 62.19 & 0.14 & 0.63 x 0.40 \\
Per-emb 35 A/B & 21.47 & 0.11 & 0.63 x 0.40 \\
SVS 13B & 140.67 & 0.36 & 0.63 x 0.40 \\
Per-emb 15 & 5.32 & 0.15 & 0.70 x 0.56 \\

Per-emb 50 & 82.65 & 0.19 & 0.63 x 0.40 \\
Per-emb 18 & 65.08 & 0.71 & 0.63 x 0.40 \\
Per-emb 37 & 11.06 & 0.09 & 0.63 x 0.40 \\
Per-emb 60 & 1.76 & 0.09 & 0.63 x 0.40 \\
EDJ2009-172 & 12.94 & 0.09 & 0.63 x 0.40 \\
Per-emb 36 & 112.16 & 0.24 & 0.63 x 0.40 \\
\end{tabular}
\end{table}

\begin{table}[h!]
\centering
\caption{Line Parameters}
\begin{tabular}{c|cccc|cccc}
\multicolumn{1}{c}{} & \multicolumn{4}{c}{C$_2$H} & \multicolumn{4}{c}{SO} \\
\hline
& \multicolumn{2}{c}{Moment 0} & Channel & & \multicolumn{2}{c}{Moment 0} & Channel & \\
Source & Max & RMS &  RMS & Beam & Max & RMS & RMS & Beam \\
& \multicolumn{2}{c}{(mJy/beam*km/s)} & (mJy/beam) & (") & \multicolumn{2}{c}{(mJy/beam*km/s)} & (mJy/beam) & (") \\
\hline
Per-emb 53 & 91.45 & 6.42 & 4.87 & 0.56 x 0.37 & 776.73 & 6.44 & 5.38 & 0.57 x 0.38 \\
Per-emb 16 & 56.44 & 5.69 & 4.73 & 0.55 x 0.37 & 81.05 & 10.07 & 4.79 & 0.67 x 0.40 \\
Per-emb 8 & 86.34 & 7.47 & 4.80 & 0.55 x 0.37 & 142.25 & 8.50 & 4.67 & 0.58 x 0.39 \\
Per-emb 11 A & 59.55 & 6.28 & 4.88 & 0.55 x 0.37 & 245.98 & 10.03 & 5.31 & 0.58 x 0.39 \\
Per-emb 1 & 51.10 & 5.86 & 4.66 & 0.55 x 0.38 & 288.86 & 24.42 & 10.25 & 0.58 x 0.39 \\
Per-emb 5 & 50.28 & 5.85 & 4.76 & 0.53 x 0.38 & 33.77 & 5.71 & 4.75 & 0.54 x 0.38 \\
Per-emb 2 & 52.62 & 7.44 & 4.68 & 0.54 x 0.38 & 179.54 & 6.48 & 4.64 & 0.54 x 0.38 \\
Per-emb 40 & 36.88 & 7.61 & 4.84 & 0.54 x 0.38 & 379.63 & 21.95 & 4.65 & 0.54 x 0.38 \\
Per-emb 10 & 83.97 & 7.33 & 4.78 & 0.54 x 0.38 & 360.86 & 7.10 & 4.74 & 0.54 x 0.38 \\
Per-emb 29 & 45.50 & 7.11 & 4.79 & 0.54 x 0.38 & 435.62 & 13.24 & 5.80 & 0.54 x 0.38 \\
Per-emb-41 N/S & 32.11 & 6.37 & 4.72 & 0.54 x 0.38 & 103.74 & 6.11 & 4.74 & 0.54 x 0.38 \\
L1448 NW & 88.36 & 7.01 & 5.32 & 0.68 x 0.43 & 411.02 & 7.67 & 5.26 & 0.67 x 0.43 \\
Per-emb 33 A & 197.92 & 8.57 & 5.35 & 0.68 x 0.43 & 168.93 & 9.04 & 5.48 & 0.67 x 0.43 \\
Per-emb 26 & 66.70 & 5.66 & 5.08 & 0.67 x 0.43 & 339.26 & 23.78 & 6.23 & 0.67 x 0.43 \\
Per-emb 22 A & 150.10 & 7.25 & 5.12 & 0.68 x 0.43 & 601.36 & 12.42 & 5.04 & 0.67 x 0.43 \\

Per-emb 25 & 63.27 & 7.85 & 5.13 & 0.70 x 0.45 & 538.26 & 6.54 & 5.02 & 0.67 x 0.43 \\
Per-emb 17 & 75.88 & 7.86 & 5.21 & 0.68 x 0.43 & 2656.63 & 59.39 & 6.36 & 0.66 x 0.43 \\
Per-emb 20 & 114.13 & 7.29 & 5.20 & 0.68 x 0.43 & 465.58 & 7.67 & 5.09 & 0.66 x 0.43 \\
L1455 IRS 2 & 30.84 & 7.10 & 5.18 & 0.67 x 0.43 & 31.28 & 6.84 & 4.99 & 0.66 x 0.43 \\
Per-emb 44 & 147.33 & 13.19 & 5.67 & 0.71 x 0.45 & 3794.34 & 15.61 & 5.42 & 0.71 x 0.45 \\
Per-emb 12 & 127.98 & 8.34 & 5.65 & 0.71 x 0.45 & 1184.92 & 51.15 & 8.75 & 0.71 x 0.45 \\
Per-emb 13 & 113.12 & 12.51 & 5.69 & 0.71 x 0.45 & 473.99 & 24.11 & 5.95 & 0.71 x 0.45 \\
Per-emb 27 & 60.35 & 8.66 & 5.78 & 0.71 x 0.45 & 3705.70 & 138.48 & 8.31 & 0.71 x 0.45 \\
Per-emb 54 & 78.32 & 9.37 & 5.82 & 0.71 x 0.45 & 224.24 & 8.54 & 5.64 & 0.71 x 0.45 \\
Per-emb 21 & 73.45 & 11.57 & 5.68 & 0.71 x 0.45 & 441.03 & 9.14 & 5.50 & 0.71 x 0.45 \\
Per-emb 14 & 111.01 & 7.53 & 5.68 & 0.71 x 0.45 & 530.19 & 7.26 & 5.42 & 0.71 x 0.45 \\
Per-emb 35 A/B & 168.85 & 7.42 & 5.68 & 0.71 x 0.45 & 1335.69 & 13.84 & 5.43 & 0.71 x 0.45 \\
SVS 13B & 149.13 & 7.72 & 5.53 & 0.71 x 0.45 & 1149.74 & 18.95 & 5.56 & 0.71 x 0.45 \\
Per-emb 15 & 35.58 & 7.43 & 5.64 & 0.71 x 0.45 & 51.32 & 10.21 & 5.51 & 0.71 x 0.45 \\

Per-emb 50 & 35.92 & 7.71 & 5.68 & 0.71 x 0.45 & 2270.27 & 46.94 & 6.28 & 0.71 x 0.45 \\
Per-emb 18 & 159.19 & 7.21 & 5.62 & 0.71 x 0.45 & 1280.34 & 15.02 & 5.41 & 0.72 x 0.45 \\
Per-emb 37 & 38.87 & 6.94 & 5.64 & 0.72 x 0.45 & 194.68 & 9.68 & 5.56 & 0.71 x 0.45 \\
Per-emb 60 & 33.61 & 9.13 & 5.81 & 0.72 x 0.45 & 51.14 & 9.94 & 5.65 & 0.71 x 0.45 \\
EDJ2009-172 & 37.73 & 8.92 & 5.67 & 0.71 x 0.45 & 43.44 & 9.95 & 5.50 & 0.72 x 0.45 \\
Per-emb 36 & 230.23 & 8.17 & 5.71 & 0.71 x 0.45 & 1031.61 & 23.11 & 5.28 & 0.71 x 0.45 \\
\hline
\end{tabular}
\end{table}
\clearpage

\begin{table}[h!]
\centering
\caption{Fitted C$_2$H Column Densities}
\begin{tabular}{c|cc}
& \multicolumn{2}{c}{Log N$_T$ (cm$^{-2})$} \\
Source & Median & Range \\
\hline
Per-emb 53 & 14.47 & 1.34 \\
Per-emb 16 & 14.36 & 1.16 \\
Per-emb 8 & 14.59 & 1.51 \\
Per-emb 11 & 14.48 & 1.21 \\
Per-emb 1 & 14.36 & 1.05 \\
Per-emb 5 & 14.28 & 1.49 \\
Per-emb 2 & 14.42 & 1.05 \\
Per-emb 40 & 14.29 & 1.00 \\
Per-emb 10 & 14.42 & 2.03 \\
Per-emb 29 & 14.11 & 1.32 \\
Per-emb 41 & 14.18 & 0.92 \\
L1448 NW & 14.48 & 2.97 \\
Per-emb 33 & 14.59 & 1.52 \\
Per-emb 26 & 14.15 & 1.69 \\
Per-emb 22 & 14.25 & 1.10 \\
Per-emb 25 & 14.21 & 1.13 \\
Per-emb 17 & 14.19 & 1.08 \\
Per-emb 20 & 14.16 & 1.29 \\
L1455 IRS 2 & 14.05 & 1.29 \\
Per-emb 44 & 14.19 & 1.11 \\
Per-emb 12 & 13.99 & 1.35 \\
Per-emb 13 & 14.04 & 1.39 \\
Per-emb 27 & 14.14 & 1.23 \\
Per-emb 54 & 14.31 & 1.42 \\
Per-emb 21 & 14.64 & 1.86 \\
Per-emb 14 & 14.08 & 0.84 \\
Per-emb 35 & 14.23 & 1.05 \\
SVS 13B & 14.49 & 1.86 \\
Per-emb 15 & 13.97 & 1.15 \\
Per-emb 50 & 14.01 & 0.91 \\
Per-emb 18 & 14.27 & 2.18 \\
Per-emb 37 & 14.00 & 0.39 \\
Per-emb 36 & 14.32 & 1.07 \\
\hline
\end{tabular}
\label{tab:column densities}
\end{table}

\bibliography{CCH_Draft}{}
\bibliographystyle{aasjournal}

\end{document}